\documentclass[12pt]{article}
\usepackage[a4paper, total={6in, 8in}]{geometry}
\usepackage{authblk}
\usepackage[utf8]{inputenc}
\usepackage{graphicx}
\usepackage{comment}
\usepackage{booktabs}
\usepackage{xcolor}
\usepackage{color}
\usepackage{multirow}
\usepackage{graphicx}

\begin{document}
%
\title{Layer Ensembles: A Single-Pass Uncertainty Estimation in Deep Learning for Segmentation}
%

%
\author[1]{Kaisar Kushibar\thanks{Corresponding author: kaisar.kushibar@ub.edu}}
\author[1]{Víctor Manuel Campello}
\author[1]{Lidia Garrucho Moras}
\author[1]{Akis Linardos}
\author[1]{Petia Radeva}
\author[1]{Karim Lekadir}
\affil[1]{University of Barcelona, Department of Mathematics and Computer Science, Barcelona, 08007, Spain.}
\date{}                     
\setcounter{Maxaffil}{0}
\renewcommand\Affilfont{\itshape\small}
\maketitle              
\begin{abstract}
Uncertainty estimation in deep learning has become a leading research field in medical image analysis due to the need for safe utilisation of AI algorithms in clinical practice. Most approaches for uncertainty estimation require sampling the network weights multiple times during testing or training multiple networks. This leads to higher training and testing costs in terms of time and computational resources. In this paper, we propose Layer Ensembles, a novel uncertainty estimation method that uses a single network and requires only a single pass to estimate predictive uncertainty of a network. Moreover, we introduce an image-level uncertainty metric, which is more beneficial for segmentation tasks compared to the commonly used pixel-wise metrics such as entropy and variance. We evaluate our approach on 2D and 3D, binary and multi-class medical image segmentation tasks. Our method shows competitive results with state-of-the-art Deep Ensembles, requiring only a single network and a single pass.

\end{abstract}

\section{Introduction}
\label{sec:introduction}
Despite the success of Deep Learning (DL) methods in medical image analysis, their black-box nature makes it more challenging to gain trust from both clinicians and patients \cite{young2021patient}.
Modern DL approaches are unreliable when encountered with new situations,
where a DL model silently fails or produces an overconfident wrong prediction.
Uncertainty estimation can overcome these common pitfalls, increasing the reliability of models by assessing the certainty of their prediction and alerting their user about potentially erroneous reports.


Several methods in the literature address uncertainty estimation in DL \cite{abdar2021review}. General approaches include: 1) Monte-Carlo Dropout (MCDropout) \cite{gal2016dropout}, which requires several forward passes with enabled dropout layers in the network during test time; 2) Bayesian Neural Networks (BNN) \cite{cinelli2021bayesian} that directly represent network weights as probability distributions; and 3) Deep Ensembles (DE) \cite{lakshminarayanan2017simple} which combines the outputs of several networks to produce uncertainty estimates. MCDropout and BNN have been argued to be unreliable in real world datasets \cite{liu2021peril}. Nonetheless, MCDropout is one of the most commonly used methods, often favourable when carefully tuned \cite{gal2016uncertainty}. Despite their inefficiency in terms of memory and time, evidence showed DE is the most reliable uncertainty estimation method \cite{beluch2018power,abdar2021review}.

There have been successful attempts that minimised the cost of training of the original DE. For example, snapshot-ensembles \cite{huang2017snapshot} train a single network until convergence, and further train beyond that point, storing the model weights per additional epoch to obtain $M$ different models. In doing so, the training time is reduced drastically. However, multiple networks need to be stored and the testing remains the same as in DE. Additionally, the deep sub-ensembles \cite{valdenegro2019deep} method uses $M$ segmentation heads on top of a single model. This method is particularly similar to our proposal -- Layer Ensembles (LE). However, LE, in contrast to existing methods, exhibits the following benefits: 
\begin{itemize}
    \item Scalable, intuitive, simple to train and test. The number of additional parameters is small compared to BNN approaches that double the number of parameters;
    \item Single network compared to the state-of-the-art DE;
    \item Unlike multi-pass BNN and MCDropout approaches, uncertainties can be calculated using a single forward pass, which would greatly benefit real-time applications;
    \item Produces global (image-level) as well as pixel-wise uncertainty measures; 
    \item Allows estimating difficulty of a target sample for segmentation (example difficulty) that could be used to detect outliers;
    \item Similar performance to DE regarding accuracy and confidence calibration.
\end{itemize}

\section{Methodology}
\label{sec:methods}
Our method is inspired by the state-of-the-art DE \cite{lakshminarayanan2017simple} for uncertainty estimation as well as a more recent work \cite{baldock2021deep} that estimates example difficulty through prediction depth. In this section, we provide a detailed explanation of how our LE method differs from other works \cite{lakshminarayanan2017simple,baldock2021deep}, taking the best from both concepts and introducing a novel method for uncertainty estimation in DL. Furthermore, we introduce how LE can be used to obtain a single image-level uncertainty metric that is more useful for segmentation tasks compared to the commonly used pixel-wise variance, entropy, and mutual information (MI) metrics.
\begin{figure}[tb]
    \centering
    \includegraphics[width=0.9\textwidth]{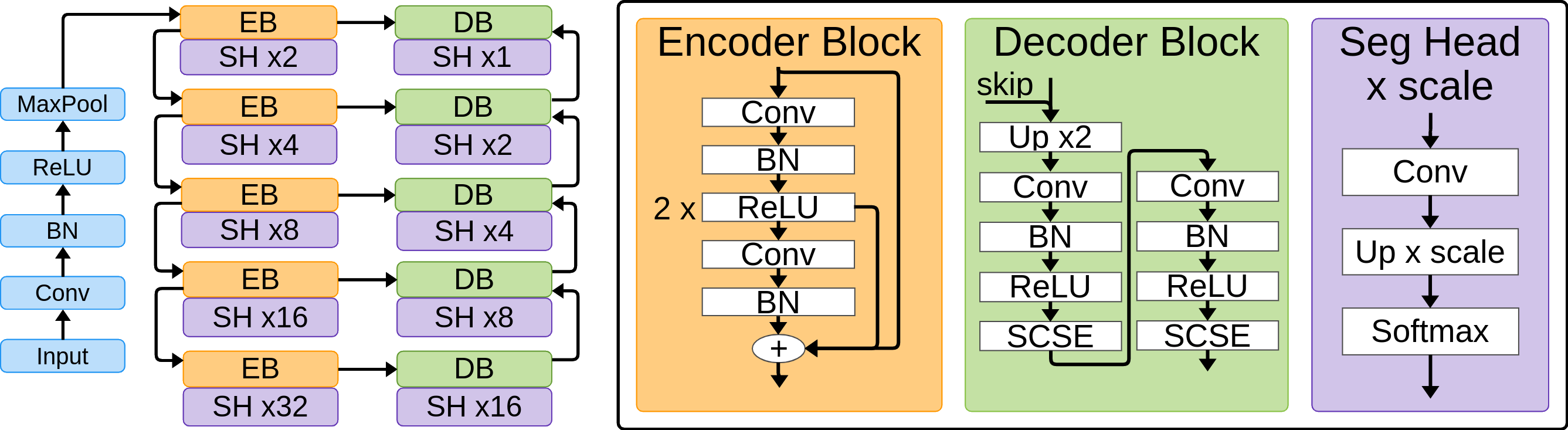}
    \caption{LE built on top of U-Net like architecture. Encoder Block (EB) in orange and Decoder Block (DB) in green have internal structures as depicted in the boxes below with corresponding colours. Ten Segmentation Heads (SH) are attached after each layer output with an up-scaling factor depending on the depth of the layer. SCSE - Squeeze and Excitation attention module. BN - Batch Normalisation.}
    \label{fig:architecture}
\end{figure}
\subsection{Prediction depth}
\label{subsec:preddepth}
Prediction Depth (PD) \cite{baldock2021deep} measures example difficulty by training k-NN classifiers using feature maps after each layer.
Given a network with $N$ layers, the PD for an input image $x$ is $L \sim [0, N]$ if the k-NN prediction for the $L^{th}$ layer is different to layer at $L-1$ and the same for all posterior layer predictions.
The authors demonstrated that easy samples have small PD, whereas difficult ones have high PD by linking the known phenomena in DL that early layers converge faster \cite{morcos2018insights} and networks learn easy data first \cite{toneva2018an}. Using PD for estimating example difficulty is appealing, however, it requires training additional classifiers on top of a pre-trained network. Moreover, using the traditional Machine Learning classifiers (e.g. k-NN) for a segmentation task is not trivial.

We extend the idea of PD to a more efficient segmentation method. Instead of k-NN classifiers, we attach a segmentation head after each layer output in the network as shown in Figure~\ref{fig:architecture}. We use a CNN following the U-Net \cite{ronneberger2015u} architecture with different modules in the decoder and encoder blocks. Specifically, we use residual connections \cite{he2016deep} in the encoder and squeeze-and-excite attention \cite{hu2018squeeze} modules in the decoder blocks. Our approach is architecture agnostic and the choice of U-Net was due to its wide use and high performance on different medical image segmentation tasks.

\subsection{Ensembles of networks of different depths}
DE has been used widely in the literature for predictive uncertainty estimation. The original method assumes a collection of $M$ networks with different initialisation trained with the same data. Then, the outputs of each of these $M$ models can be used to extract uncertainty measurements (e.g. variance). As we have shown in Figure~\ref{fig:architecture}, ten segmentation heads were added after each layer. Then, LE is a compound of $M$ sub-networks of different depths. Since each of the segmentation heads is randomly initialised, it is sufficient to cause each of the sub-networks to make partially independent errors \cite{Goodfellow-et-al-2016}. The outputs from each of the segmentation heads can then be combined to produce final segmentation and estimate the uncertainties, similarly to DE. Hence, LE can be considered equivalent to DE, but using only one network model.

\subsection{Layer agreement as an image-level uncertainty metric}
\begin{figure}[tb]
    \centering
    \includegraphics[width=0.9\textwidth]{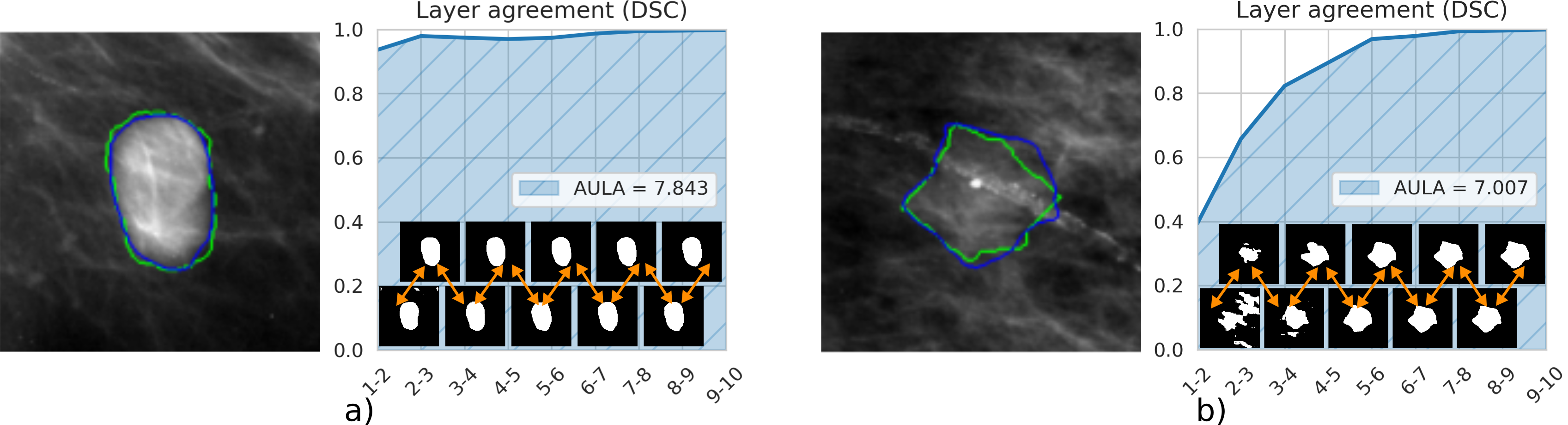}
    \caption{Layer Agreement curve. a) A high contrast lesion: large AULA and low uncertainty. b) A low contrast lesion and calcification pathology is present: small AULA and higher uncertainty. Arrows represent the correspondence between layers 1 and 2, 2 and 3, etc. DSC -- Dice Similarity Coefficient. Green contours are ground truths.}
    \label{fig:aula_example}
\end{figure}

As we have stated above, LE is a combination of sub-networks of different depths. It can also be viewed as stacked networks where the parameters of a network $f_t$ is shared by $f_{t+1}$ for all $t \in \mathopen[0, N\mathclose)$, where $N$ is the total number of outputs. This sequential connection of $N$ sub-networks allows us to observe the progression of segmentation through the outputs of each segmentation head. We can measure the agreement between the adjacent layer outputs -- e.g. using the Dice coefficient -- to obtain a layer agreement curve. Depending on the network uncertainty, the agreement between layers will be low, especially in the early layers (Figure~\ref{fig:aula_example}). We propose to use the Area Under Layer Agreement curve (AULA) as an image-level uncertainty metric. In the following sections, we demonstrate that AULA is a good uncertainty measure to detect poor segmentation quality, both in binary and multi-class problems.

\section{Materials and Implementation}
\label{sec:materials_and_implementation}
We evaluate our proposal on two active medical image segmentation tasks: 1) Breast mass for binary 2D; and 2) Cardiac MRI for multi-class 3D segmentation. We assess LE in terms of segmentation accuracy, segmentation quality control using AULA metric, and example difficulty estimation using PD. For all the experiments, except for the example difficulty, LE is compared against the state-of-the-art DE approach and also a plain network without uncertainty estimation (referred as Plain).

\subsection{Datasets}
We use two publicly available datasets for the selected segmentation problems. Breast Cancer Digital Repository (BCDR) \cite{moura2013benchmarking} contains 886 MedioLateral Oblique (MLO) and CranioCaudal (CC) view mammogram images of 394 patients with manual segmentation masks for masses. Original images have a matrix size of $3328 \times 4084$ or $2560 \times 3328$ pixels (unknown resolution). We crop and re-sample all masses to patches of $256 \times 256$ pixels with masses centred in the middle, as done in common practice \cite{rezaei2021review}. We randomly split the BCDR dataset into train (576), validation (134), and test (176) sets so that images from the same patient are always in the same set.

For cardiac segmentation, the M\&Ms challenge (MnM) \cite{campello2021multi} dataset is utilised. We use the same split as in the original challenge -- 175 training, 40 validation, and 160 testing.
All the images come annotated at the End-Diastolic (ED) and End-Systolic (ES) phases for the Left Ventricle (LV), MYOcardium (MYO), and Right Ventricle (RV) heart structures in the short-axis view. In our experiments, both time-points are evaluated together. All MRI scans are kept in their original in-plane resolution varying from isotropic $0.85 mm$ to $1.45 mm$ and slice-thickness varying from $0.92 mm$ to $10 mm$. We crop the images to $128 \times 128 \times 10$ dimensions so that the heart structures are centred.

\subsection{Training}
The same training routine is used for all the experiments, with only exception in batch-size: 10 for breast mass and 1 for cardiac structure segmentation. The network is trained for 200 epochs using the Adam optimiser to minimise the generalised Dice \cite{sudre2017generalised} and Cross-Entropy (CE) losses for breast mass and cardiac structure segmentation, respectively. For the multi-class segmentation, CE is weighted by $0.1$ for background and $0.3$ for each cardiac structure. An initial learning rate of $0.001$ is set with a decay by a factor of $0.5$ when the validation loss reaches a plateau. Common data augmentations are applied including random flip, rotation, and random swap of mini-patches of size $10 \times 10$. Images are normalised to have zero-mean and unit standard deviation. A single NVIDIA GeForce RTX 2080 GPU with 8GB of memory is used. The source code with dependencies, training, and evaluation is publicly available\footnote{Github link will be presented soon.}.

\subsection{Evaluation}
Testing is done using the weights that give the best validation loss during training. The final segmentation masks are obtained by averaging the outputs of individual networks in DE ($M = 5$). For LE, we tried both averaging the sub-network outputs and using the well-known Simultaneous Truth And Performance Level Estimation (STAPLE) algorithm \cite{warfield2004simultaneous} that uses weighted voting. Both results were similar and for brevity we present only the version using STAPLE.

We evaluate LE and DE using the common uncertainty metrics in the literature -- the pixel-wise variance, entropy, and MI, and they are summed for all pixels/voxels in cases where an image-level uncertainty is required. The AULA metric is used for LE. The network confidence calibration is evaluated using the Negative Log-Likelihood metric (NLL). It is a standard measure of a probabilistic model’s quality that penalises wrong predictions that have small uncertainty \cite{quinonero2005evaluating}. Note that AULA can also be calculated by skipping some of the initial segmentation heads.

\section{Results}
\subsection{Segmentation performance and confidence calibration}
Table~\ref{tab:segmentation_and_calibration} compares the segmentation performance of LE with DE and Plain models in terms of Dice Similarity Coefficient (DSC) and Modified Hausdorff Distance (MHD). Two-sided paired t-test is used to measure statistically significant differences.
\begin{table}[tb]
\caption{Segmentation and confidence calibration performance for Plain U-Net, DE, and LE on BCDR and MnM datasets. The values for DSC, MHD, and NLL are given as `mean(std)'. $\uparrow$ - higher is better, $\downarrow$ - lower is better. Best values are in bold. Statistically significant differences compared to LE are indicated by `*'.}
\label{tab:segmentation_and_calibration}
\centering
\scriptsize
\begin{tabular}{@{}lccccccc@{}}
\toprule
\multicolumn{1}{c}{} & \multicolumn{3}{c}{BCDR -- breast mass segmentation} &  & \multicolumn{3}{c}{MnM -- all structures combined} \\ \midrule
Method & DSC $\uparrow$ & MHD $\downarrow$ & NLL $\downarrow$ &  & DSC $\uparrow$ & MHD $\downarrow$ & NLL $\downarrow$ \\
Plain & *0.865(0.09) & *1.429(1.72) & *2.312(1.35) &  & 0.900(0.11) & \textbf{1.061(2.69)} & 0.182(0.41) \\
DE & 0.870(0.09) & 1.373(1.76) & *0.615(0.54) &  & *0.896(0.13) & 1.465(4.86) & *\textbf{0.157(0.33)} \\
LE & \textbf{0.872(0.084)} & \textbf{1.317(1.692)} & \textbf{0.306(0.25)} &  & \textbf{0.903(0.10)} & 1.302(5.31) & 0.173(0.37) \\ \midrule
 & \multicolumn{3}{c}{MnM -- Structure-wise DSC $\uparrow$} &  & \multicolumn{3}{c}{MnM -- Structure-wise MHD $\downarrow$} \\ \cmidrule(lr){2-4} \cmidrule(l){6-8} 
Method & LV & MYO & RV &  & LV & MYO & RV \\
Plain & 0.882(0.13) & *0.804(0.12) & *0.826(0.15) &  & \textbf{1.313(3.63)} & \textbf{1.303(2.79)} & 2.884(10.89) \\
DE & \textbf{0.885(0.14)} & *0.804(0.13) & 0.829(0.16) &  & 1.536(5.6) & 1.500(3.98) & \textbf{2.113(5.67)} \\
LE & 0.883(0.13) & \textbf{0.809(0.11)} & \textbf{0.832(0.14)} & \textbf{} & 1.525(5.86) & 1.529(5.50) & 2.525(8.92) \\ \bottomrule
\end{tabular}
\end{table}
In breast mass segmentation, LE performs similarly to DE and Plain model for both DSC and MHD metrics. The NLL of LE, however, is significantly better compared to others $(p < 0.001)$. For cardiac structure segmentation, the combined DSCs for all methods are similar and MHD of Plain is slightly better. NLL of DE ($0.157\pm0.33$) is significantly better than ours ($0.173\pm0.37$) $(p < 0.05)$, however, LE can achieve an NLL of $0.140\pm0.23$ by skipping less layers without compromising segmentation performance (see Figure~\ref{fig:segmentation_quality_and_layer_skipping}, right). In our experiments, skipping the first three and five outputs gave the best results in terms of correlation between uncertainty metrics and segmentation performance for breast mass and cardiac structure tasks, respectively (see Table~\ref{tab:correlation}). Skipping all but the last segmentation head in LE is equivalent to the Plain network. In terms of structure-wise DSC, all methods are similar for all the structures. Plain method has a slightly better MHD compared to DE and LE for the LV and MYO structures $(p > 0.05)$, and DE is better for the RV structure compared to LE $p > 0.05$.

The ranking across all metrics in Table~\ref{tab:segmentation_and_calibration} are: LE (1.58), DE (2.08), and Plain (2.33), showing that on average LE is better. Overall, segmentation performance of all three are similar and LE has a better confidence calibration.

\subsection{Segmentation quality control}
\begin{figure}[tb]
    \centering
    \includegraphics[width=0.9\textwidth]{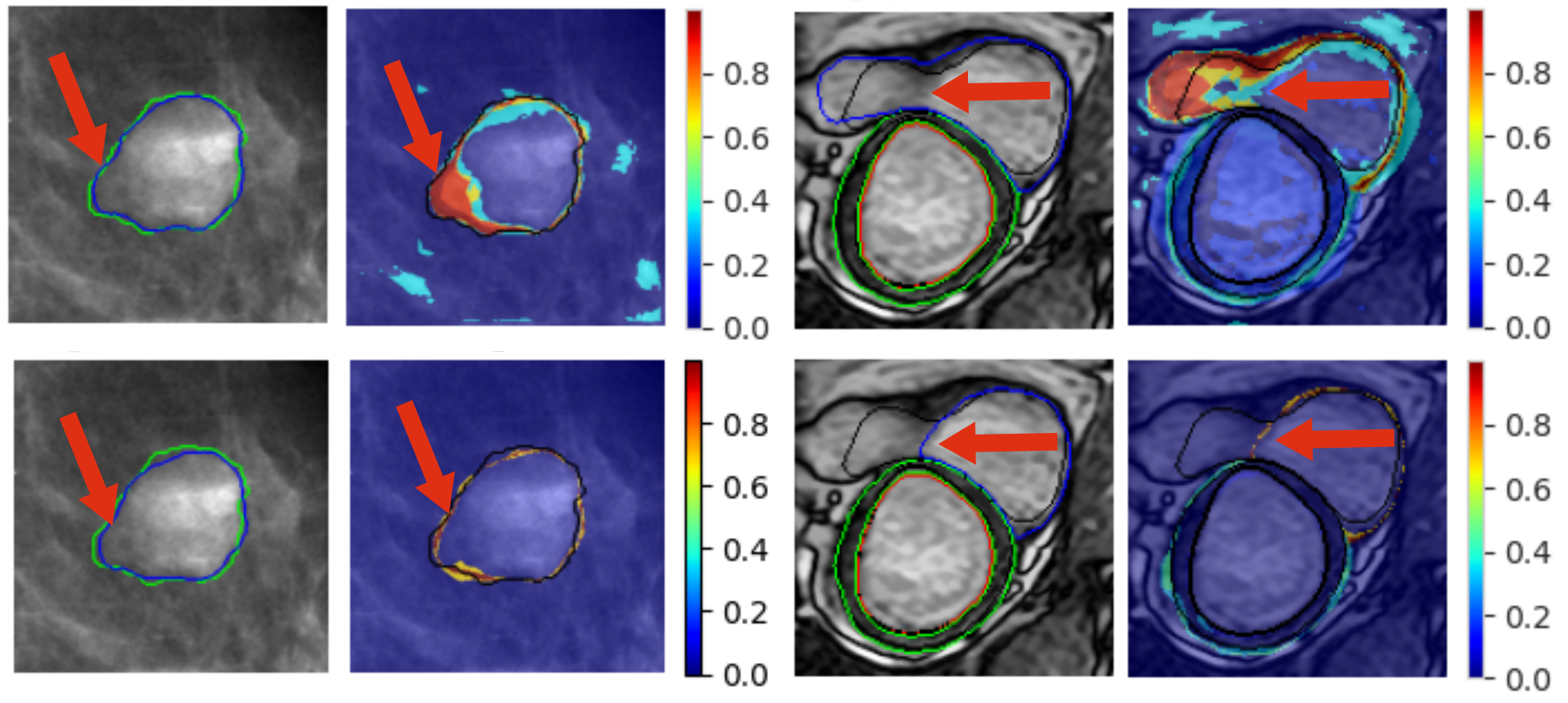}
    \caption{Examples of visual uncertainty heatmaps based on variance for high uncertainty areas (red arrows) using LE (top) and DE (bottom) for breast mass and cardiac structure segmentation. Black and green contours correspond to ground truth.}
    \label{fig:qualitative_examples}
\end{figure}

\begin{table}[tb]
\caption{Spearman's correlation of segmentation metrics with uncertainty metrics for breast mass and cardiac segmentation tasks. Absolute highest values are shown in bold.}
\label{tab:correlation}
\centering
\scriptsize
\begin{tabular}{@{}rccccccccc@{}}
\toprule
\multicolumn{1}{c}{} & \multicolumn{4}{c}{BCDR} &  & \multicolumn{4}{c}{MnM} \\ \midrule
\multicolumn{1}{c}{} & Entropy & MI & Variance & AULA &  & Entropy & MI & Variance & AULA \\
DE-DSC & -0.783 & -0.785 & \textbf{-0.785} & N/A &  & -0.323 & -0.433 & -0.377 & N/A \\
LE-DSC & 0.615 & 0.597 & 0.620 & \textbf{0.785} &  & 0.221 & 0.207 & 0.203 & \textbf{0.649} \\
DE-MHD & 0.762 & \textbf{0.764} & 0.763 & N/A &  & 0.401 & 0.499 & 0.447 & N/A \\
LE-MHD & -0.594 & -0.575 & -0.598 & -0.730 &  & -0.309 & -0.313 & -0.300 & \textbf{-0.571} \\ \bottomrule
\end{tabular}%
\end{table}
\begin{figure}
    \centering
    \includegraphics[width=0.9\textwidth]{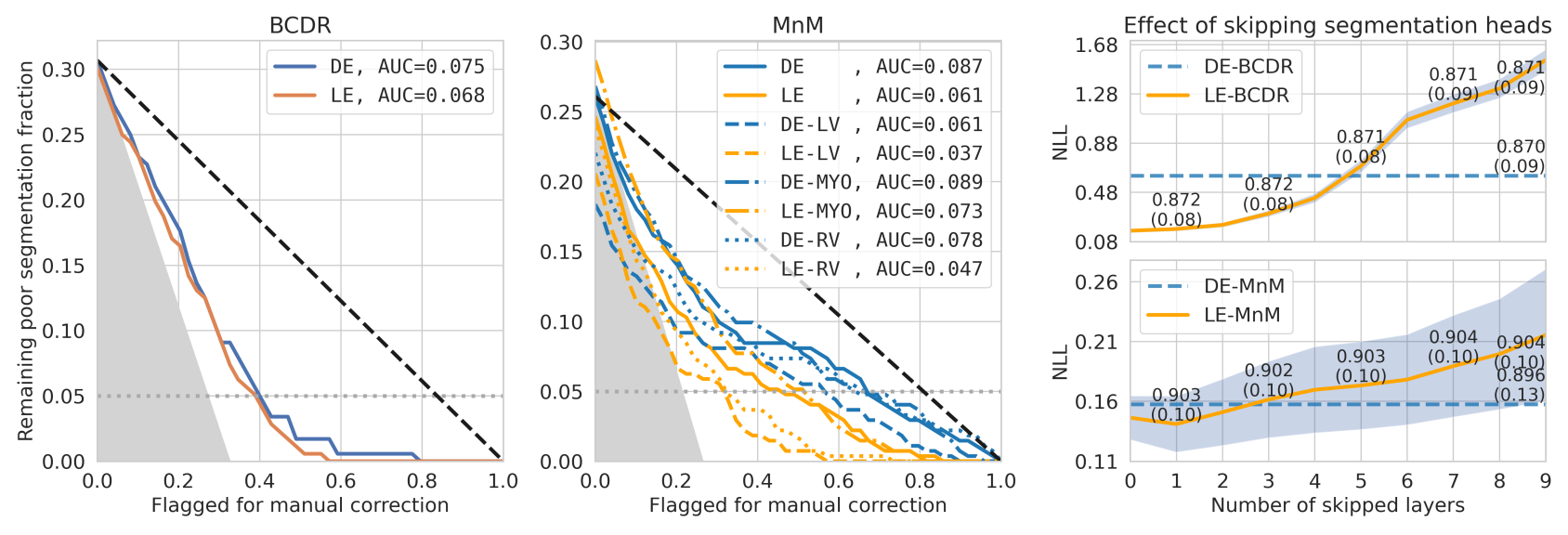}
    \caption{\textbf{Segmentation quality control} for DE and LE. The following are averaged indicators for: random flagging (dashed black); remaining 5\% of poor segmentations (dotted grey); and ideal line (grey shaded area). \textbf{The effect of skipping initial segmentation head outputs} on model calibration. Numbers on top of the lines represent DSC in `mean(std)' format. Shaded areas are standard deviations for NLL.}
    \label{fig:segmentation_quality_and_layer_skipping}
\end{figure}
We evaluate our uncertainty estimation proposal for segmentation quality control, similarly to \cite{ng2018estimating}, and compare it to the state-of-the-art DE. We use the proposed AULA uncertainty metric to detect poor segmentation masks and the variance metric for DE. Figure~\ref{fig:segmentation_quality_and_layer_skipping} shows the fraction of remaining images with poor segmentation after a fraction of poor quality segmentation images are flagged for manual correction. We consider DSCs below $0.90$ as poor quality for both segmentation tasks.
We set the threshold for the cardiac structure following the inter-operator agreement identified in \cite{bai2018automated}, and use the same value on the threshold for masses.
As proposed in \cite{ng2018estimating}, the areas under these curves can be used to compare different methods. It can be seen that LE and DE are similar in terms of detecting poor quality segmentations, with LE achieving slightly better AUC for all the cases -- mass, combined and structure-wise cardiac segmentations. Table~\ref{tab:correlation} supports this statement by confirming high correlation between AULA and segmentation metrics. In BCDR, both are somewhat close to the averaged ideal line. For the cardiac structure segmentation, all the curves take a steep decline, initially being also close to the averaged ideal line indicating that severe cases are detected faster. Moreover, as can be seen in Figure~\ref{fig:qualitative_examples}, DE's uncertainty maps are overconfident, while LE manages to highlight the difficult areas. We believe that having such meaningful heatmaps is more helpful for the clinicians (e.g. for manual correction). More visual examples including entropy and MI are given in Appendix.

\subsection{Example difficulty estimation}
\begin{figure}[tb]
    \centering
    \includegraphics[width=0.9\textwidth]{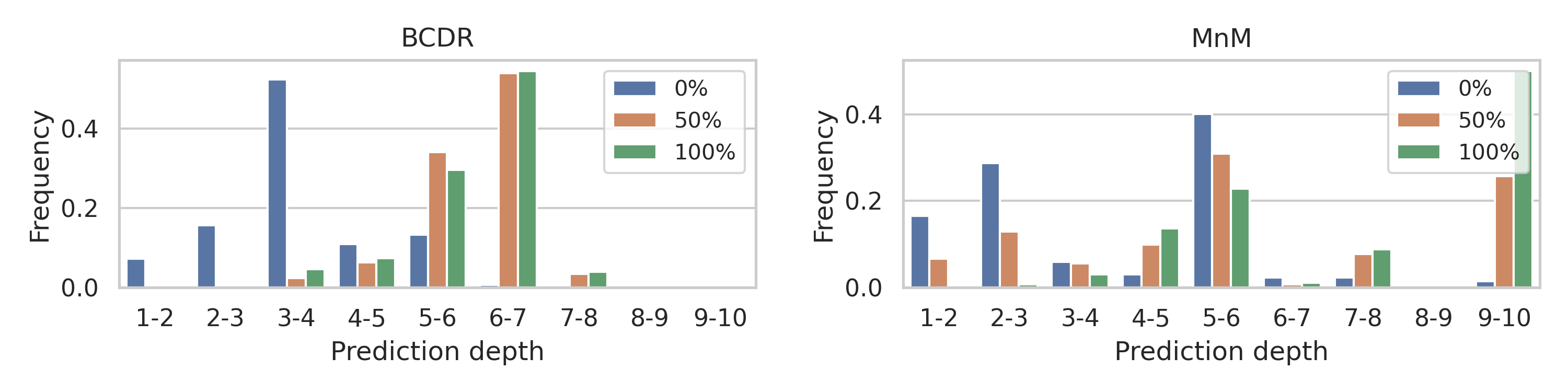}
    \caption{PD distribution with 0\%, 50\%, and 100\% of the images corrupted by Gaussian noise -- MnM $\mathcal{N}(0.3, 0.7)$, BCDR Random Convolutions with kernel-size (37, 37). Layer agreement threshold is 0.90 in terms of DSC for both datasets.}
    \label{fig:prediction_depth}
\end{figure}
We evaluate example difficult estimation using PD by perturbing the proportion of images in the test set. We added random Gaussian noise to MnM dataset and used Random Convolutions \cite{xu2021robust} in BCDR as the model was robust to noise in mammogram images. Examples of perturbed images are provided in Appendix. Then, for a given sample, PD is the largest $L$ corresponding to one of the $N$ segmentation heads in a network, where the agreement between $L$ and $L-1$ is smaller than a threshold that is the same as in segmentation quality control. In this sense, PD represents the minimum number of layers after which the network reaches to a consensus segmentation. Figure~\ref{fig:prediction_depth} shows how the  distribution of PD shifts towards higher values as the number of corrupted images increases in the test set for both BCDR and MnM datasets.
Overall, cardiac structure segmentation is more difficult than breast mass segmentation while the latter is more robust to noise.
This demonstrates how PD can be used to evaluate example difficulty for the segmentation task and detect outliers.

\section{Conclusions}
\label{sec:conclusions}
We proposed a novel uncertainty estimation approach that exhibits competitive results to the state-of-the-art DE method using only a single network. Compared to DE, our approach produces a more meaningful uncertainty heatmaps and allows estimating example difficulty in a single pass.
Experimental results showed the effectiveness of the proposed AULA metric to measure an image-level uncertainty measure. The capabilities of both AULA and PD were demonstrated in segmentation and image quality control experiments. We believe that the efficient and reliable uncertainty estimation that LE demonstrates will pave the way for more trustworthy DL applications in healthcare.





\section{Acknowledgements}
This study has received funding from the European Union’s Horizon 2020 research and innovation programme under grant agreement No 952103.

%
%
%
\bibliographystyle{splncs04}
\bibliography{bibliography}

\newpage
\section*{Appendix}
\subsection*{Corrupted image examples to increase prediction depth}
\begin{figure}[h]
    \centering
    \includegraphics[width=0.9\textwidth]{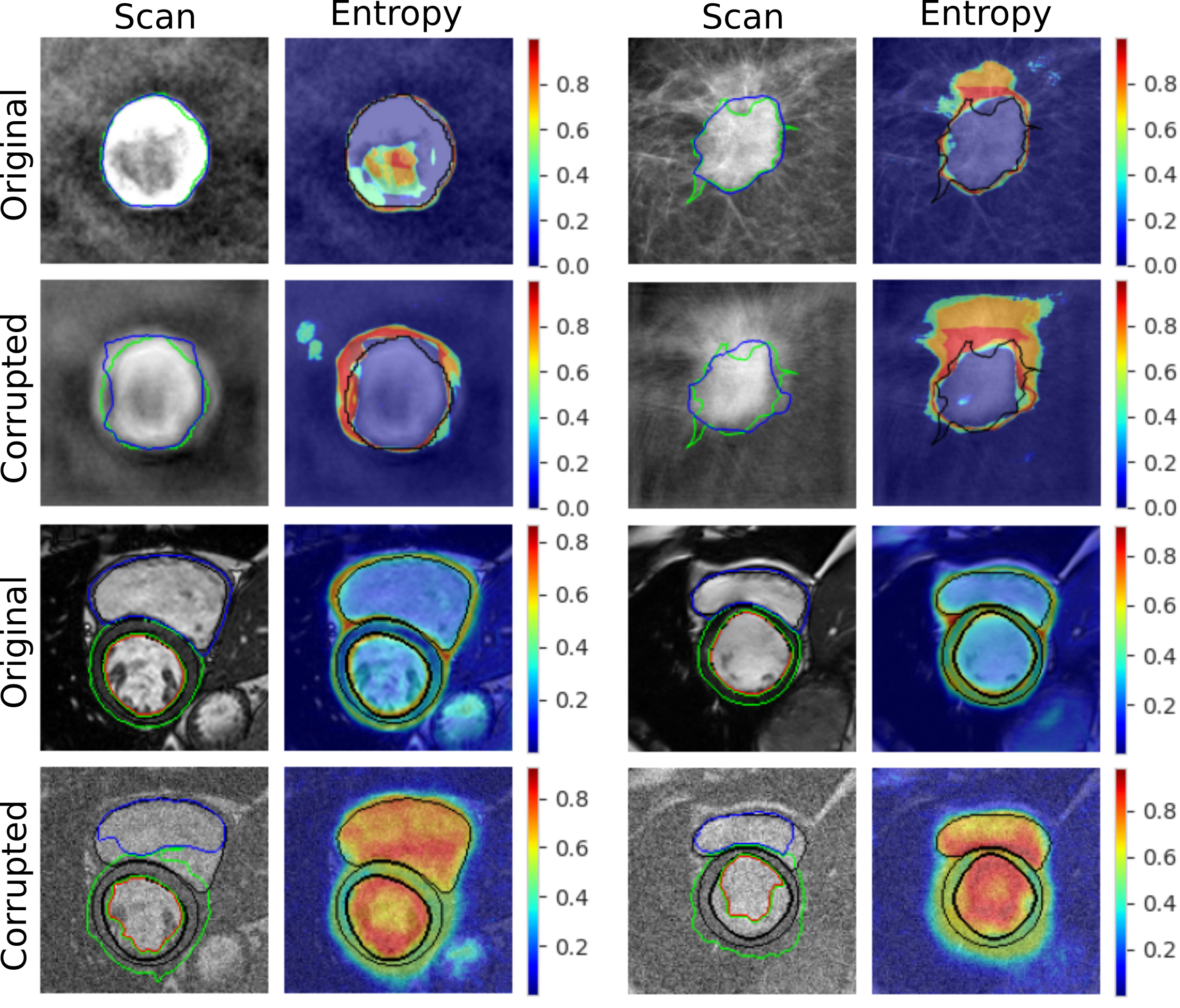}
    \caption{Examples of corrupted images and their effect on the entropy map. Black and green contours correspond to ground truth.}
    \label{fig:corrupted images}
\end{figure}
\newpage
\subsection*{Qualitative examples}
\label{sec:qual_examples}
\begin{figure}[h]
    \centering
    \includegraphics[width=0.9\textwidth,trim={0 15.3cm 0 0},clip]{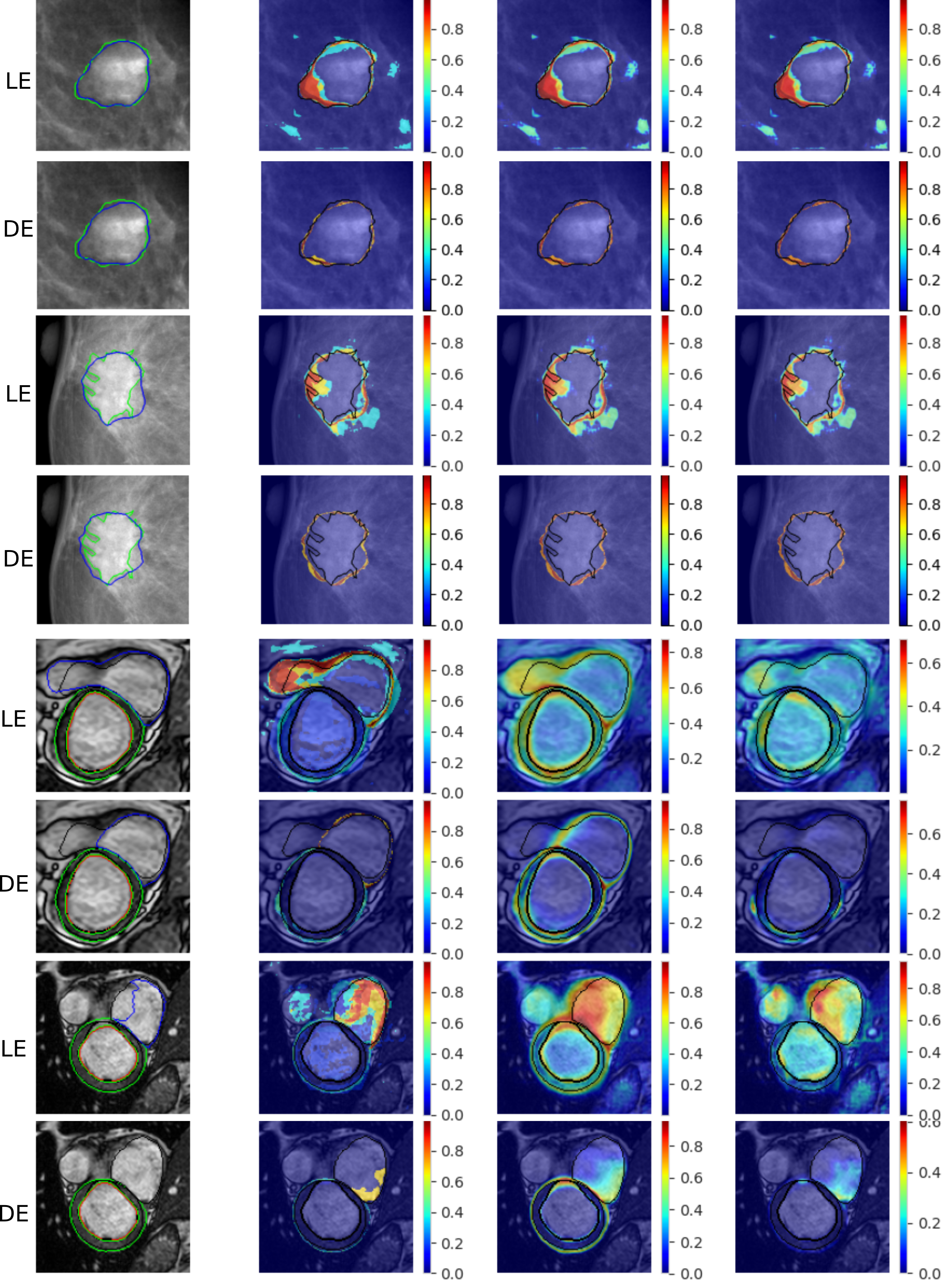}
    \caption{\textbf{BCDR}. Examples of visual uncertainty heatmaps based on variance, entropy, and mutual information. Black and green contours correspond to ground truth.}
    \label{fig:more_examples1}
\end{figure}
\begin{figure}[h]
    \centering
    \includegraphics[width=0.9\textwidth,trim={0 0 0 15.3cm},clip]{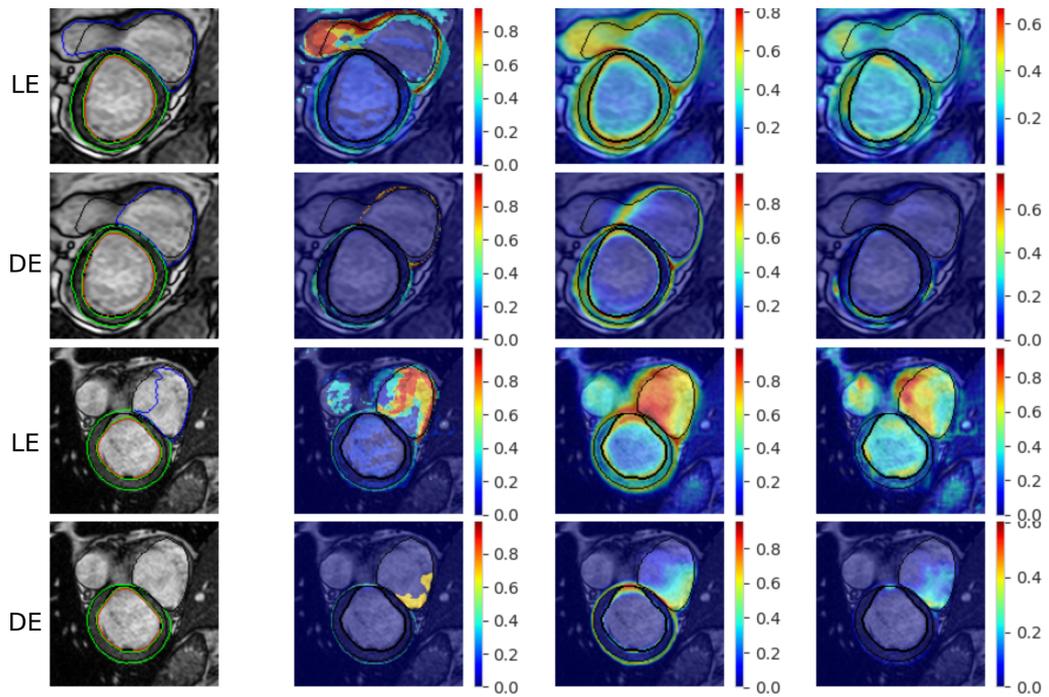}
    \caption{\textbf{MnM}. Examples of visual uncertainty heatmaps based on variance, entropy, and mutual information. Black contours correspond to ground truth.}
    \label{fig:more_examples2}
\end{figure}
%




\end{document}